\documentclass[manuscript=article]{achemso}
\usepackage[version=3]{mhchem} 
\usepackage{multirow}
\usepackage{natmove}
\usepackage{bm}				
\usepackage{color}
\usepackage{ulem}

\author{Rodrigo Ramos}
\affiliation{Instituto de F{\'\i}sica, Universidade de S\~ao Paulo, 05508-900 S\~ao Paulo, SP, Brazil}
\affiliation{Now at: Centro Universit{\'a}rio das Faculdades Metropolitanas Unidas, S\~ao Paulo, SP, Brazil}
\author{Mar{\'\i}lia J. Caldas}
\affiliation{Instituto de F{\'\i}sica, Universidade de S\~ao Paulo, S\~ao Paulo, SP, Brazil}
\email{mjcaldas@usp.br}
\title{P3HT-Fullerene Blends: a Classical Molecular Dynamics Simulation}
\begin{document}
\begin{abstract}

Blends of polymer and C$_{60}$-derived molecules are in the spotlight in recent years for application in organic photovoltaics, forming what is known as bulk-heterojunction active layers. The character of the heterojunction is determinant, with clear relevance of morphology and phase separation.
To better understand the morphology of the systems, we present a classical molecular dynamics (CMD) simulation of polymer/fullerene (P3HT/C$_{60}$) blends, coming from different starting points, using the specifically designed Nanomol Force Field based on the Universal Force Field. We use not-so-short regioregular polymers with 30 hexyl-thiophene units ($\sim$5kg$\cdot$mol$^{-1}$ molecular weight) and, adopting a $\sim$ 1:1 mass proportion for the polymer:molecule blend, we simulate cells of $\sim$ 40 thousand atoms, at room temperature and normal pressure conditions. We find that, independently of the starting-point spatial distribution of C$_{60}$ molecules relative to P3HT chains, segregated or isotropic, the fullerene molecules show a tendency to segregate, and phase separation is the dominating regime.
\end{abstract}
\section*{TOC Graphic}
\begin{figure}%
\centering
\includegraphics[width=5.1cm]{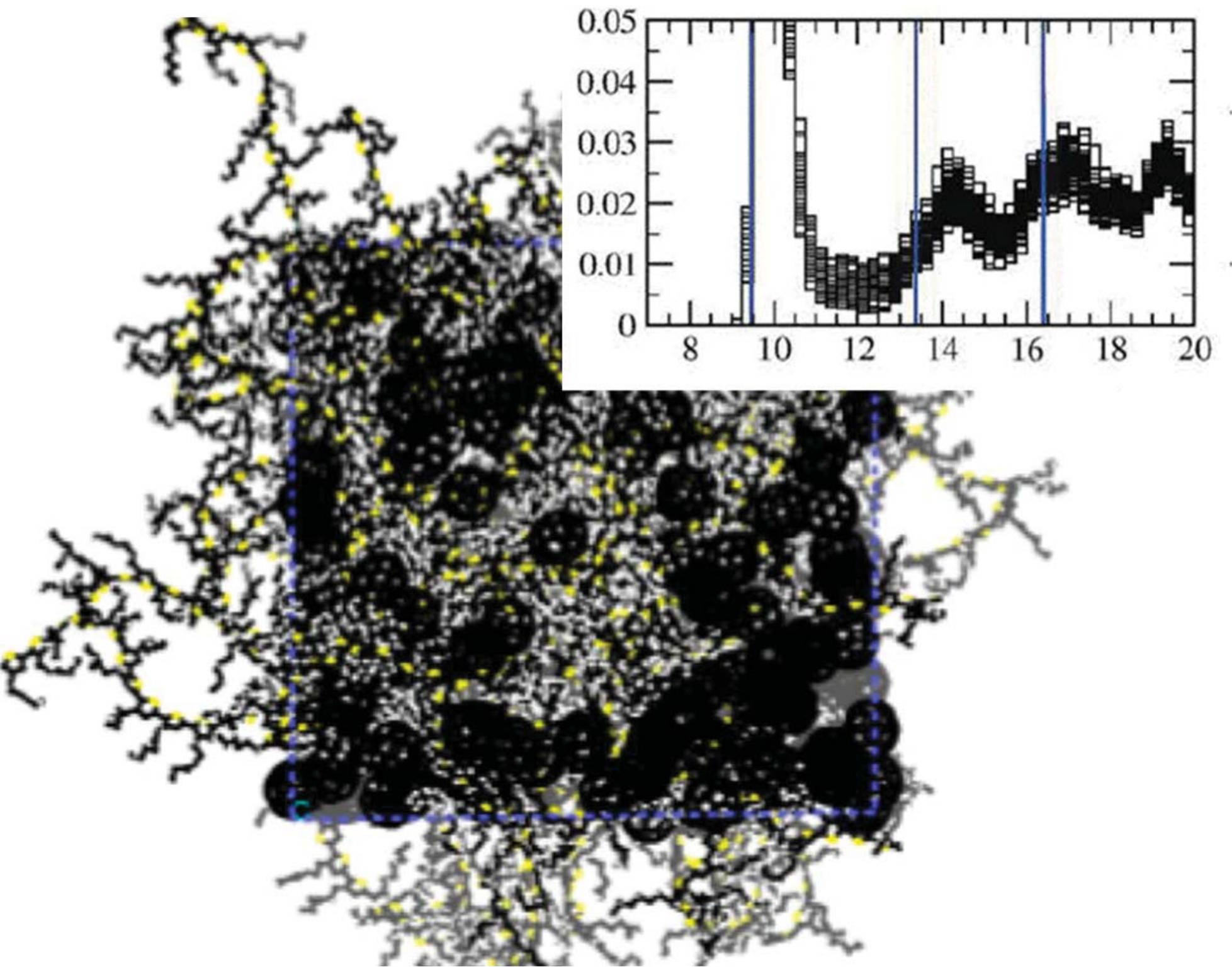}
\end{figure}

\textbf{Keywords}: Thiophene, P3HT, fullerene, blend, C$_{60}$, Classical Molecular Dynamics, UFF
\newpage
Organic photovoltaics is a rich and growing area of research, and among the most promising device architectures we find those in which the active layer contains thiophene polymers and fullerenes, more specifically poly(3-hexyl-thiophene) P3HT and [6,6]-Phenyl-C61-butyric acid methyl ester PCBM.\cite{brab+10am,verp+10afm,dang+11am,chen+11nl,holm+13semsc,chin+16nsc,busb+17acsami,rich+17cr}  This active layer is known as a bulk heterojunction (BHJ), in the sense that we will find regions (domains) where there is prevalence of P3HT mass proportion, others where the prevalence is of PCBM. The nature of the frontier between domains can be well-defined (bicontinuous heterojunction) or not, as well as the prevalence inside a domain, that can be absolute (bicontinuous) or not, in the sense that the material, in domains or in the frontier, is formed by a blend of the two components. There is still ongoing discussion on these topics, however it is agreed that the photo-conversion efficiency is related to charge transfer, exciton migration and recombination, internally in the domains and in the frontier between domains; a plethora of experimental and theoretical works concerning these issues is available, and it emerges that one of the most important features regards the blend morphology.\cite{schi+05cm,guo+10jacs} While PCBM is a reasonably simple molecule, consisting of a fullerene C$_{60}$ molecule with a methyl-ethyl ester side chain, due to the lower cost of C$_{60}$ production, blends consisting of P3HT and C$_{60}$ (or a mixture of the two molecules) are also interesting\cite{rait+07semsc,li+09oe,tada-onod12semsc,comi+15mcp,mori+17jpst}. Since the alkyl side chains can strongly interact, the morphology of the final complete blend, polymer and molecule, can be more or less complex, more homogeneous or phase-separated, depending on the average length of the P3HT main chain. The molecules can segregate and form molecular almost crystalline domains, or migrate into the polymer region, depending wether the vicinal polymer region is more or less crystalline.

P3HT belongs to the second generation of electronic polymers\cite{brab+10am}, and is one of the most studied to this day\cite{busb+17acsami}. The hexyl functionalization enhances solubility and, when regioregular, may contribute\cite{nort07prb,guo+10jacs} to the formation of stacked crystalline domains depending on the growth method; it is however usually found that in devices the polymer is present in complex morphology, with important proportion of amorphous domains. As a semiconducting polymer, the optical and transport properties are defined by the character and distribution of electron and hole states in the bulk, which in turn are defined by the \textit{local} morphology. The exciton splitting and charge transfer properties when considering the blend are also defined by local morphology, and the investigation of such transitions, theory or experiment, must take that into account. In order to study theoretically such properties we must have access to a reliable description of the structural phases, which is only accessible through Classical Molecular Dynamics (CMD) -- since now we are not describing periodic crystals, with a number of atoms-per-cell that could be studied through atomistic first-principles calculations \cite{buss+02apl,ruin+03sm,ferr+04prb,humm+05pssb}.

Concerning CMD calculations, it can be done through simplified formulations, usually referred to as coarse-grained molecular dynamics, which allows for treatment of systems with a very large number of atoms; this procedure has been recently applied to P3HT and blends\cite{negi+16mts,root+16mm,ales+17jacs,yoo+17cms}, and contributes to the understanding of morphology in these complex organic compounds. On the other hand, when we need to look more deeply into the local structural characteristics of these compounds, we must go to atomistic simulations which, even in the classical formulation\cite{luko+17ms}, restricts the number of included atoms to some tens of thousands.

We present here an atomistic CMD simulation of condensed P3HT, and P3HT blends not with PCBM but with simple fullerene C$_{60}$; the models are built with not-so-short oligomers for the thiophene unit (30 3HT mers) and with P3HT:C$_{60}$ blend mass proportion very close to 1:1, as often found\cite{dang+11am,holm+13semsc,busb+17acsami} in the experimental literature for P3HT:PCBM blends. We apply for this a well-tuned force-field, which we describe explicitly, based on the Universal Force Field of Rapp\'{e} and coll.\cite{rapp+92jacs} We start from two different initial models, random or phase separated P3HT/C$_{60}$ spatial distribution. We find that after reaching equilibrium conditions, the P3HT chains present mostly non-linear structure, and concomitantly the polymer domains present disordered amorphous character. We see weak intermixing of C$_{60}$ and P3HT for the phase-separated, and strong clustering of C$_{60}$ molecules for the random starting point.  Our main conclusions are that for blends at room temperature, even for these non-crystalline polymer domains, phase separation is dominating.

\section{Methodology}

One of the most used and reliable force-fields for CMD of polymers is the Universal Force Field UFF of Rapp\'{e} and collaborators\cite{rapp+92jacs}, which however presents unusual problems  in the case of thiophenes. Specifically, it was found\cite{alve-cald09sm} that the standard charge assignment scheme of UFF (charge equilibration) for sulfur S atoms results in negative partial charge $Q_S<0$, as frequently happening in other molecules, which is not correct for thiophene (T) since the special [$=HC-S-CH=$] sequence in the T rings results in a positive $Q_S>0$ effective charge. Furthermore, the dihedral ring-to-ring torsion angle in oligothiophenes (OTs) was also found not to be properly described, which lead us to adapt the related parameters.

Here we describe only briefly the methodology for reparametrization of the force field. In short, the atom-atom interaction in the UFF can be broadly separated in bonded and non-bonded functions. For bonded-functions we have 2, 3 and 4-body potentials: bond-length (2), direct angle (3), dihedral and inversion angles (4). For non-bonded functions the FFs normally adopt pair-potentials including electrostatic Coulomb long-range interactions, van der Waals attraction and Pauli repulsion; these last two are grouped in the Lennard-Jones (LJ) format in the UFF. Adequate LJ parameters are  crucial for the reliable description of the condensate morphology.

Starting with the non-bonded potentials, as mentioned above we need reliable values for atomic charges. We will adopt fixed atomic charges and define specific atomic types also for the C- and H-atoms in the T-ring and side-chains.  To arrive at the atomic type charges ATC (more than one ATC for the same atom type of the UFF) we perform calculations with Density Functional Theory (DFT), using the FHI-aims code\cite{blum+09cpc} with the PBE exchange-correlation functional\cite{perd+96prl}, known to provide reliable structural properties for organic compounds. The partial charges for all atoms in our chosen prototype-molecules ensemble were calculated through the Hirschfeld method\cite{hirs77tca}  and averaged in order to achieve the series of ATCs. For the Lennard-Jones potentials we calculate the dispersion coefficients $D_{IJ}$ and equilibrium distance $R_{IJ}$ for atom pairs of the ensemble using the Tkatchenko-Scheffler formalism\cite{tkat-sche09prl}. Our final values for ATCs, Ds and Rs are tabulated in the Supplementary Information.

We check our ATC results with experimental data\cite{bak+61jms,harr+53jcs,frin+77ahc,spoe-cols98cp} for the electric dipole of the single thiophene molecule T1, that range from 0.46 to 0.60D; our PBE result adopting the TS approach is 0.50D, in excellent agreement with the experimental data, and from our Nanomol-FF ATCs we arrive at 0.52D. With respect to the complete reparametrization of the dihedral T-T angle and all non-bonded parameters for the HT molecules and polymers, we selected 13 different\cite{nist} thiophene-based molecular crystals,  from clean T2 to longer oligomer chains, including also linked oligomers, alkyl-terminated oligomers and polymers (details in the Supplementary Information); for this ensemble we realized geometry optimization through Molecular Mechanics within the conjugate gradient procedure and with rigid convergence criteria. The maximal deviations in lattice constants and angles with respect to experimental data are below 10$\%$, again indicating that our parameters allow for reliable description of condensed systems.

Moving to the bonded, dihedral inter-ring torsion angle parametrization, we base our procedure on experimental data\cite{bucc+jacs74} for bi-thiophene T2, which state that at room-temperature in liquid crystalline solvent, 70$\%$ of the molecules present quasi-antiparallel arrangement with dihedral angle of 140$^{\circ}$, while the remaining 30$\%$  equilibrate at 40$^{\circ}$, quasi-parallel. This is in very good agreement with many-body theoretical calculations which allows us to adopt the full theoretical potential curve\cite{alve06phd} for the parametrization. To simulate longer polymers we perform calculations for the T4 oligomer, and adopt the displaced-dihedral form for the 4-body potential. The equation and related parameters are shown in the Supplementary Information. The other bonding 2-body parameters from the original UFF are kept.

All CMD and Molecular Mechanics (MM, temperature 0K) calculations were performed with the Cerius$^2$ package, Accelrys Inc.\cite{cerius2}. We adopt periodic boundary conditions in order to simulate condensates, with fixed number of particles N. For CMD we use fixed or variable volume V and pressure P, and fixed temperature T=300K. The time step for the sequential resolution of the MD equations is 10$^{-3}$ps, using the Verlet integrator\cite{verl67pr}. Depending on the sequence of steps, we adopt microcanonical NVE, canonical NVT  or isothermal-isobaric NPT ensembles\cite{alle-tild05book}. For NVT we use the Berendsen scheme\cite{bere+84jpc}, and for NPT the Parrinello-Rahman\cite{parr-rahm81jap}. The relaxation time is 0.1ps.

For MM we adopt different energy minimization protocols, conjugate-gradient for the crystalline structures used for definition of the Nanomol FF parameters, and the Smart Minimizer present in the Cerius$^2$ package for the condensates. This last protocol incorporates sequential steps of steepest descent, quasi-Newton and truncated-Newton minimization\cite{rapp-case97book}.

\begin{figure}[!htb]
\begin{center}
\begin{tabular}{cc}
\hline\\
\vspace{0.0cm}\\
\multicolumn{2}{c}{\includegraphics[width=0.7\textwidth]{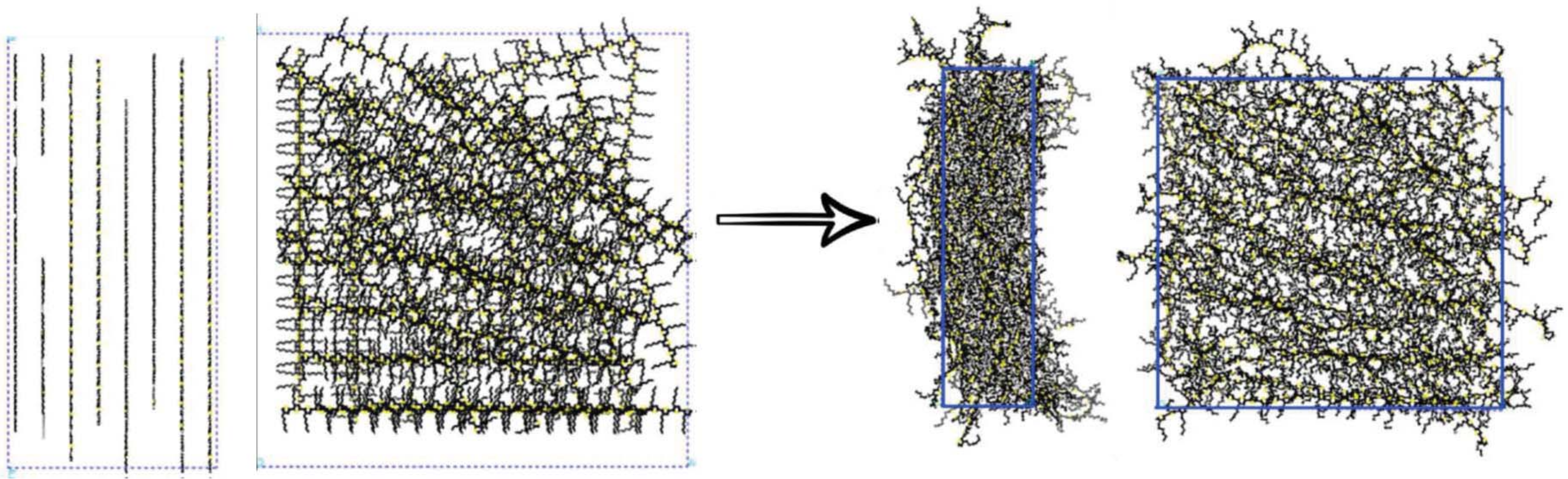}}\\
\multicolumn{2}{c}{\includegraphics[width=0.7\textwidth]{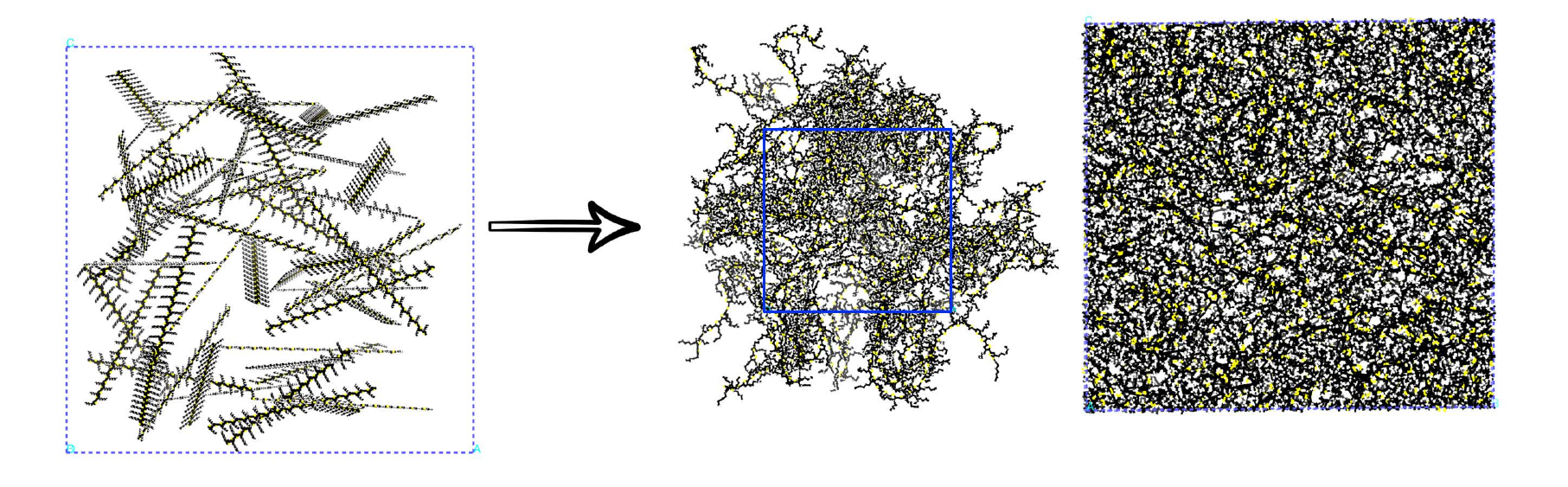}}\\
\hline\\
\vspace{0.0cm}\\
\includegraphics[width=0.35\textwidth]{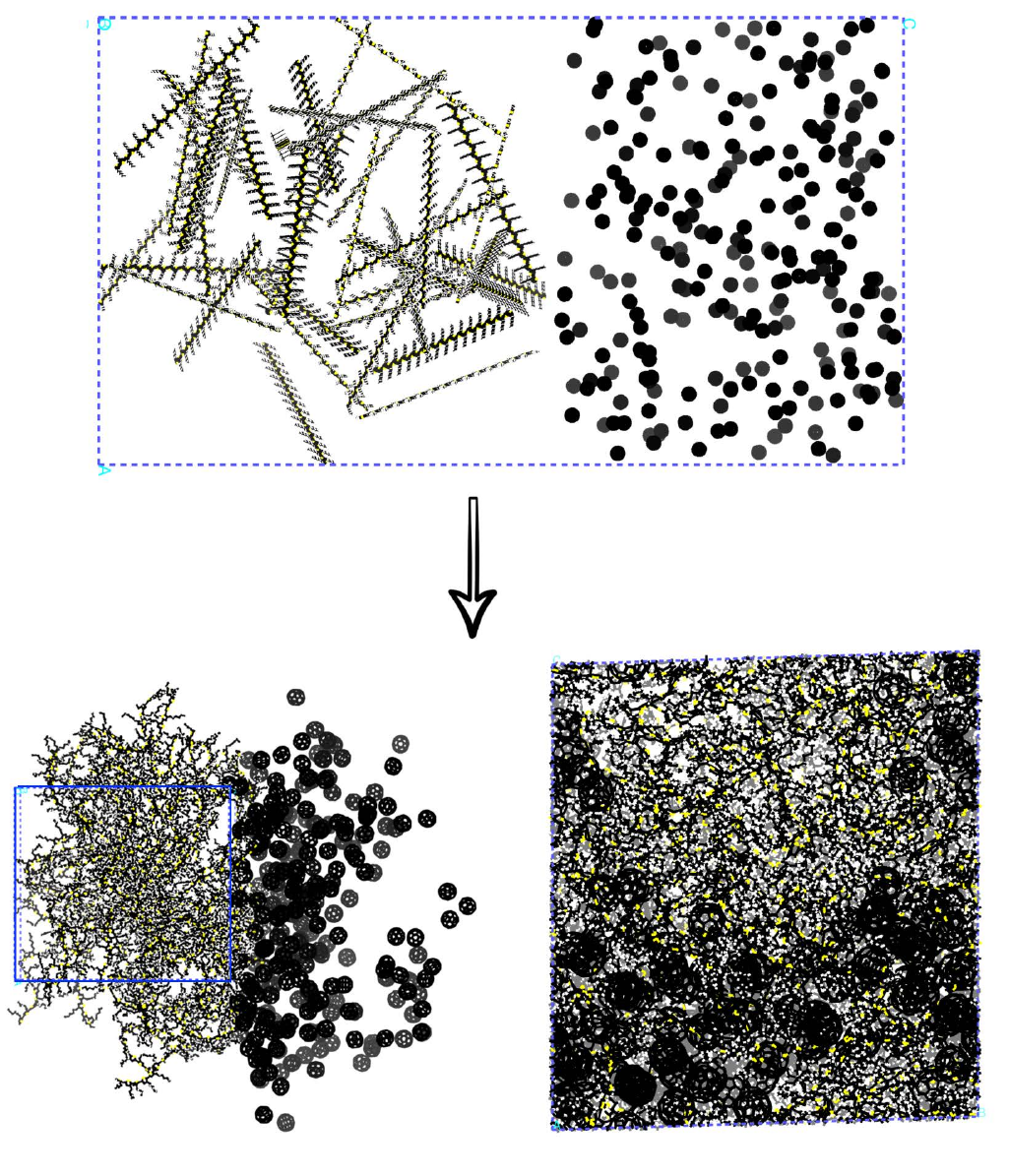} & \includegraphics[width=0.4\textwidth]{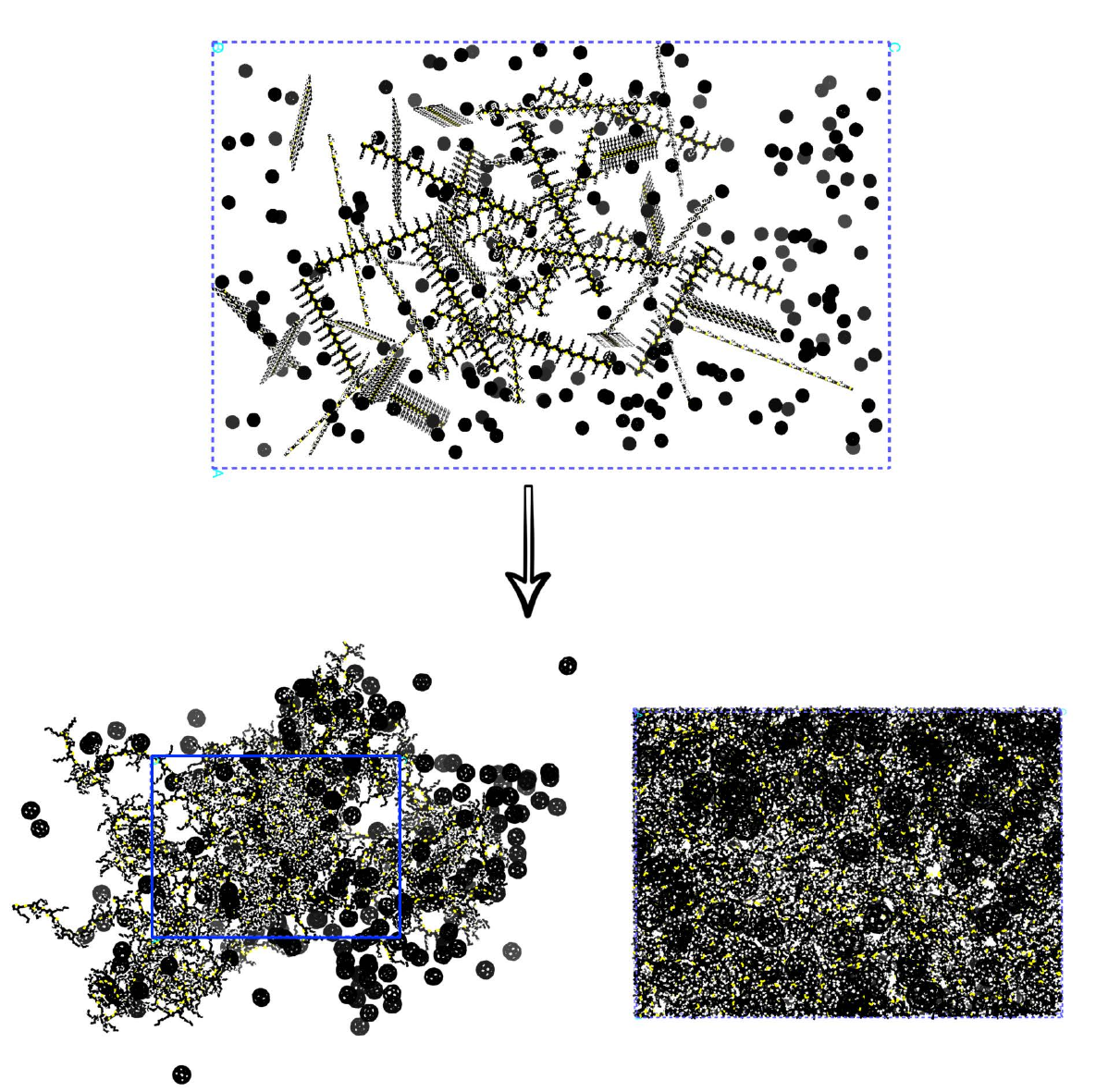}\\
\hline
\end{tabular}
\caption{Condensed amorphous models obtained in this work; we show the initial random distribution, and the final equilibrated distribution in unit-by-unit and full cell view. Top, P3HT laminar deposition, side and top view; Center, P3HT isotropic distribution;  Bottom left, segregated P3HT:C$_{60}$ blend (different viewpoints for the unit-by-unit and full cell); Bottom right, isotropic P3HT:C$_{60}$ blend. }
\label{condensation}
\end{center}
\end{figure}

We simulate condensates of regioregular P3HT, pure and in blends with fullerene C$_{60}$. We use not-so-short oligomers with 30 3HT-units O3HT$_{30}$, molecular weight $\sim$5kg$\cdot$mol$^{-1}$, feasible for our atomistic molecular simulation and small but comparable to regular experimentally-grown samples\cite{dang+11am,holm+13semsc,spol+15oe}. The simulation is performed for four different initial configurations, namely pure P3HT in laminar and isotropic distributions and P3HT:C$_{60}$ in segregated and isotropic blends. In the case of pure P3HT these two distributions simulate the first a patterned deposition, and the second a spin-cast deposition. For pure P3HT we include 40 chains of O3HT$_{30}$ for the laminar distribution (30080 atoms), and 50 chains for the isotropic (37600 atoms); for the blends we include 40 chains of O3HT$_{30}$ and 250 C$_{60}$ molecules (45080 atoms). The initial and final atomic distributions are illustrated in \ref{condensation}. The initial spatial distribution of units in done in a very disperse molecular packing, with large inter-molecular distances, via the Packmol package.\cite{packmol} A smooth, cautious procedure is then adopted in order to allow for realistic arrangement of the molecules and oligomers, without unnatural high temperature or pressure effects. We first reduce slowly, by hand, the cell parameters, and for each fixed-cell parameters perform CMD at 300K; this slow cell-reduction scheme continues until reaching the minimum of the van der Waals energy for the cell. With these optimal cell parameters we start the free NPT CMD minimization until convergence is attained.

\section{Results and Discussion}

The formation of heterojunctions in P3HT:PCBM films has been extensively investigated from the experimental side, and more than one detail of the growth method has been highlighted as a factor to be considered\cite{schi+05cm,agos+11afm,dang+11am,chen+11nl,holm+13semsc,spol+15oe,chin+16nsc,rich+17cr}. When the complete film is grown by deposition and then annealing of a in-solution mixture, as normally done, the factors can be, at least: for P3HT, molecular weight or average molecular mass (it is expected that crystalline regions will form\cite{schi+05cm} if it arrives at $\sim$ 10 kg$\cdot$mol$^{-1}$), and degree of regioregularity; for the blend, mass proportion, solvation of each compound in the deposition solution (choice of solvent), annealing time and temperature, and finally even the architecture of the monitored sample\cite{busb+17acsami}.  A general understanding is that PCBM will diffuse (with thermal annealing) only to disordered regions of P3HT, will not migrate into already crystallized regions of the polymer, and on the other hand more probably will segregate forming the junction structure. When the junction is built from bilayer deposition of the two compounds the process is different,\cite{trea+11aem} since the P3HT original layer already has a good fraction of crystalline regions; it is found that with temperature there is fast diffusion of PCBM, however only into the disordered regions of P3HT.

The adoption of C$_{60}$ for the blend is highly desirable, and experimental investigations of the P3HT:C$_{60}$ blend have been carried out. The blend properties have also been shown to depend on the growth methodology\cite{rait+07semsc,li+09oe,tada-onod12semsc,comi+15mcp,mori+17jpst} and, as expected, the behavior of C$_{60}$ is slightly different from that of PCBM, however the main trends are retrieved: dependence on the solvent, best efficiency for 1:1 mass proportion, and so forth. P3HT, C$_{60}$ or PCBM are soluble in polar solvents and usually, for the blend, polymer and molecule weights are mixed to arrive at a desired mass proportion, that can vary\cite{maye+09afm,dang+11am,geli+14sci} from 1:4 to 3:2 (P3HT:molecule), being this last possibly the maximum proportion allowing for efficient exciton charge-transfer splitting.

In this work, we use the NanomolFF to investigate phase segregation in a P3HT/C$_{60}$ blend. We first simulate pure systems, and in sequence the blends. We implement here for the blend simulations a mass proportion 11:10, very close to the suggested\cite{rait+07semsc,dang+11am,holm+13semsc} proportion 1:1.

\subsection{P3HT properties in films and blends}

We first analyse the P3HT single-chain morphological distribution for the different condensates, using for that the ring-to-ring angular distribution. For each T-ring characteristic vector axes can be defined: $\vec{c}$, linking the two C-atoms bonded to the S-atom, which defines the bonding or chain direction; a normal vector $\vec{n}$ perpendicular to the T-plane; and a basal `` dipole'' $\vec{d}$-vector, $\vec{n}=\vec{c}\times\vec{d}$ in-plane and pointing from the 2 other C-atoms to the S-atom. In this way, for each single chain we can follow the orientation of one ring to the next T$_{i+1}$-T$_{i}$, through the angles
$\varphi_{i+1,i}=\arccos({\vec{n}_{i+1}}\cdot{\vec{n}_i})$; $\theta_{i+1,i}=\arccos[(\vec{n}_{i+1}\cdot\vec{c}_i) / (\sqrt{(\vec{n}_{i+1}\cdot\vec{d}_i)^2+(\vec{n}_{i+1}\cdot\vec{n}_i)^2})]$. In this, $\varphi$ describes mostly the dihedral torsion angle between neighbor units while $\theta$ describes roughly the concavity at that segment.

\begin{figure}[!htb]
\begin{center}

\includegraphics[width=\textwidth]{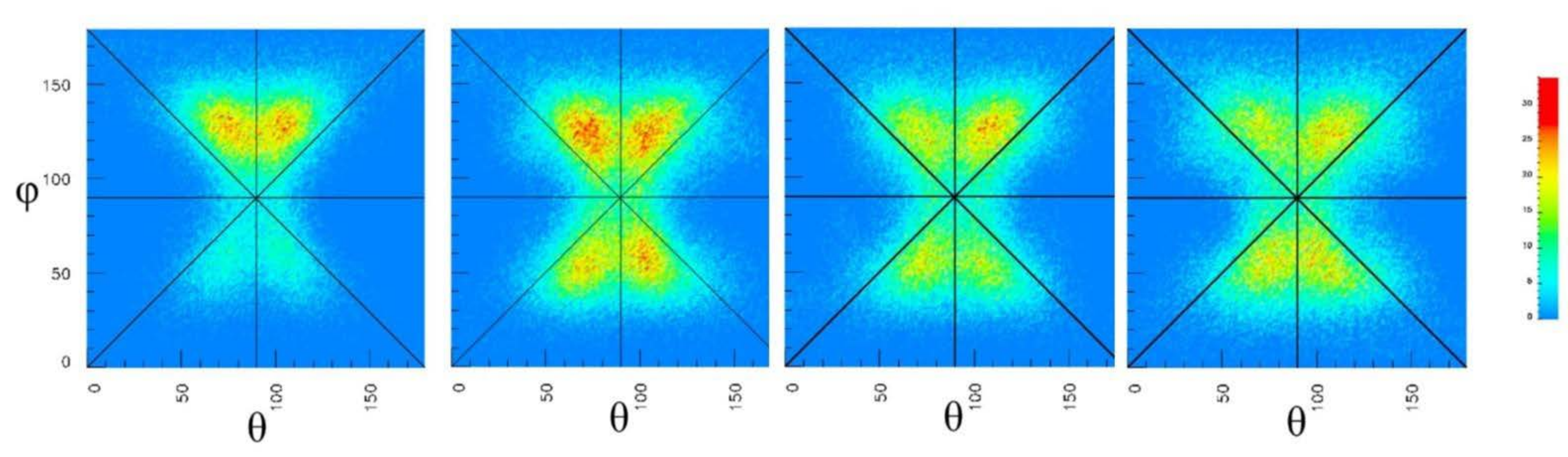}

\caption{(\textit{color online}) Map of torsion and concavity angles $(\varphi, \theta)$ for the condensates: Laminar P3HT (left); Isotropic P3HT (middle left); Segregated P3HT:C$_{60}$ (middle right); Isotropic P3HT:C$_{60}$ (right). Extreme values $\theta=0^{\circ},\ 180^{\circ}$ would correspond to extreme kinks, not realistic, moderate angles correspond to chain concavity; $\varphi<90^{\circ}$ non-alternate, $\varphi>90^{\circ}$ alternate ring-to-ring torsion angle. }
\label{map}
\end{center}
\end{figure}

We show in  \ref{map} the results for the four condensates. As a first output, we see that for the laminar-deposition condensate we mostly find, as the final geometry for the individual chains, the alternate $\varphi > 90^{\circ}$ ring-to-ring pattern, even so with a non-zero signal for $\varphi<90^\circ$, and still some remaining traces of linear (straight ideal  $\theta=90^{\circ}$) ring-to-ring orientation. For the pure P3HT isotropic condensate we see a ``butterfly'' pattern characteristic of torsion angles maxima around 130$^{\circ}$ (quasi-antiparallel), and with a smaller but definite proportion around 50$^{\circ}$ (quasi-parallel); the proportion of linear ring-to-ring orientation is almost null. For the blend simulations, for which the polymer region is isotropically initialized, we also find the butterfly pattern now with mainly the same percentage of parallel and anti-parallel ring-to-ring pattern. At the same time, we see that the full majority of chains in these three phases show concavity, that is, linearity is not a relevant characteristic, which is a clear indication of the disordered, amorphous character of the domains. We should point out here that this is expected for the molecular weight we simulate.\cite{schi+05cm}

Summarizing our results for the polymer configurations in the different simulation schemes, we see that in the laminar structure the individual chains mostly maintain linearity and ring-to-ring alternation. In the next three cases, for pure or blended P3HT, linearity is not a signature anymore, and we find a strong presence of chiral mer-to-mer angles (butterfly pattern). More significant is that for the two blend simulations, initially segregated or isotropic, the final individual polymer structures show strong similarity.

\subsection{C$60$ and P3HT-C$60$ properties in the blends}

\begin{figure}[!htb]
\begin{center}

\includegraphics[width=\textwidth]{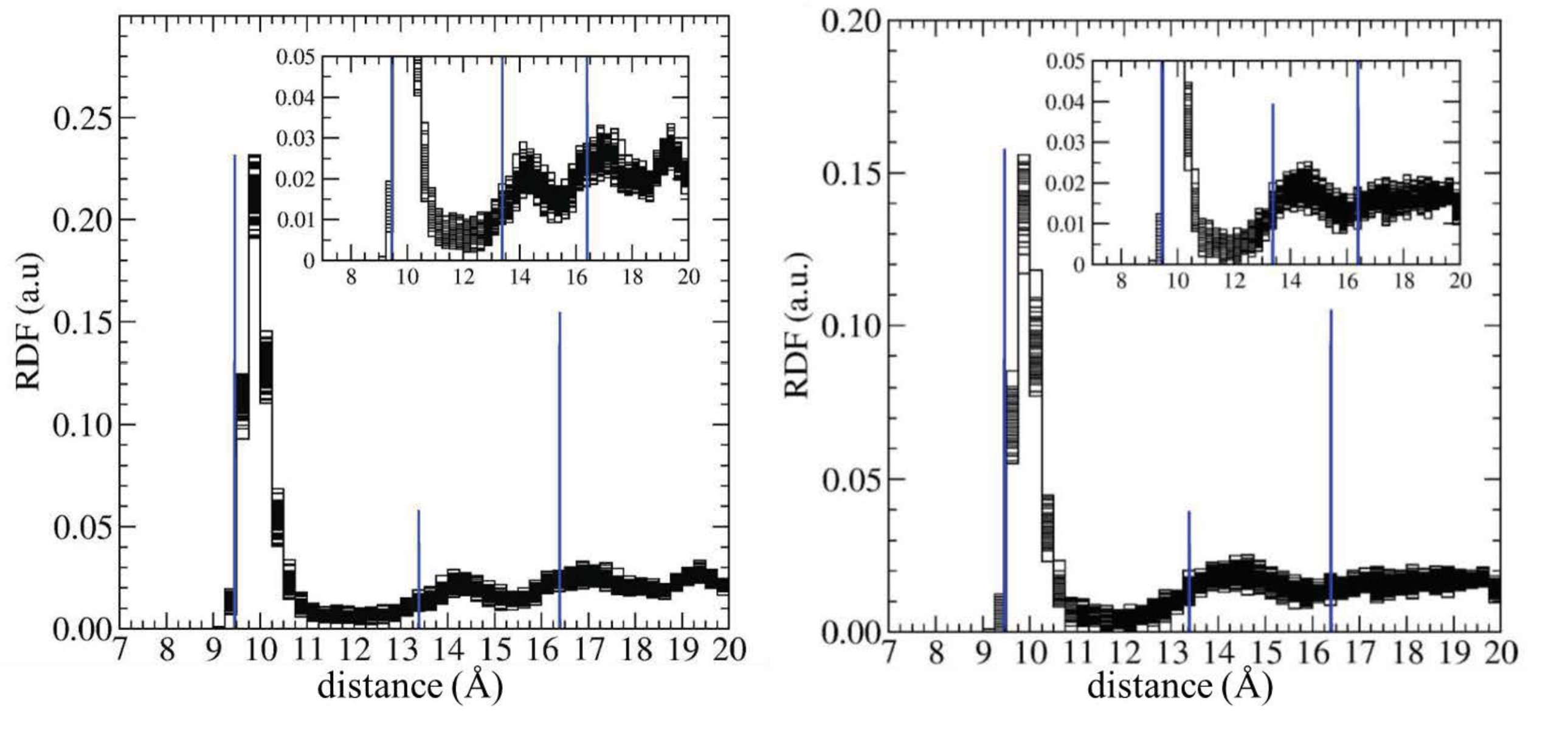}

\caption{(\textit{color online}) Radial distribution function RDF for  fullerene-fullerene molecules, 100 snapshots, in the segregated (left) and isotropic (right) blends P3HT:C$_{60}$. Arbitrary units, same normalization ratio for the two systems. The vertical blue lines correspond to first- to third-neighbor RDF signals for ideal \textit{fcc} crystalline C$_{60}$, see text, obtained with the same Nanomol-FF at 0K. }
\label{RDF-C-C}
\end{center}
\end{figure}

We now analyse the difference between the two blend models focusing on the C$_{60}$ distribution, and to do that we use the radial distribution function RDF of fullerene molecules, shown in \ref{RDF-C-C} for the two blend simulations. In the figure we show also, as (blue) vertical lines, the results obtained through Nanomol-FF at 0K (geometry minimization) for the ideal crystal, starting with the \textit{fcc} ideal lattice; we find the structure to be very close to the \textit{fcc}, with one lattice constant slightly larger than the other two, and thus the second-neighbor distances result very close but not equal, so the lines are not summed with the resolution used in the figure. Focusing on the blend results, the RDF's are  built from 100 snapshots of each CMD simulation, and we see that the first-neighbor distance is a little larger than in the 0K structure, as expected, due to temperature effects; the value is in very good accord with available experimental data\cite{heba93arm} for the pure fullerene compound at normal temperatures. For the segregated model we find that the RDF peaks are clearly defined up to the fourth neighbor distances, and specially up to the second neighbor (with a spread), which indicates the prevalence of C$_{60}$ local ordered domains. It should be noticed that also for the initially isotropic blend we see (less defined) RDF peaks to the third neighbor, which points to  a natural segregation of the C$_{60}$ molecules. The main conclusions here are that we see a tendency to formation of separated polymer and fullerene domains in both simulations, that is, even with the relatively low polymer molecular weight we adopt, that as we find does not favor formation of crystalline P3HT domains, our results indicate segregation of C$_{60}$ molecules.

\begin{figure}[!htb]
\begin{center}

\includegraphics[width=\textwidth]{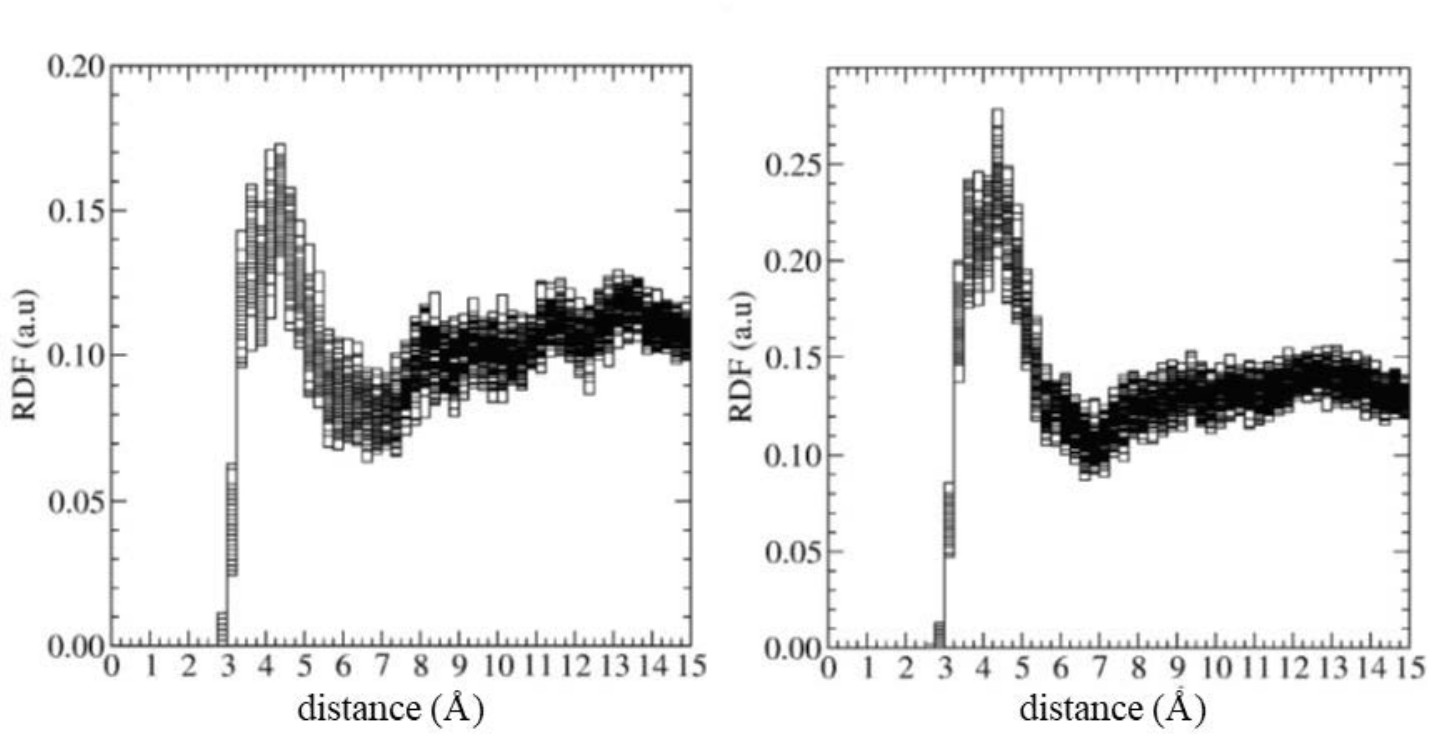}

\caption{Radial distribution function RDF for fullerene molecule-thiophene unit, 100 snapshots, in the segregated (left) and isotropic (right) blends P3HT:C$_{60}$. Arbitrary units, same normalization ratio for the two systems. The distance is measured from the center of a C$_{60}$ molecule to the center of the thiophene rings, subtracted the normal fullerene molecular radius.}
\label{RDF-T-C}
\end{center}
\end{figure}

We finally analyse the RDF, shown in \ref{RDF-T-C}, for the distance from a fullerene molecule surface to the central point of a thiophene ring. For both types of blend, we see in the figure a broad peak at $\sim$4{\AA} measuring the distance of the closest neighbor T-rings from a fullerene surface; the decrease at $\sim$5{\AA} indicates the curvature of the polymer chain when adjacent to the molecule, that is, a T-chain will not be linear when adjacent to a $C_{60}$ molecule.

We now focus on the presence of thiophene-fullerene T-C$_{60}$ compared to C$_{60}$-C$_{60}$ first-neighbor pairs from \ref{RDF-C-C} and \ref{RDF-T-C}. We note that we find a $\sim$60$\%$ higher occurrence of T-C$_{60}$ pairs in the initially isotropic compared to the initially segregated blend, and the opposite proportion for C$_{60}$-C$_{60}$ pairs: these proportions are not so high considering the difference is the initial structures, and again point to the natural segregation of $C_{60}$ molecules.

Summarizing our results, overall we see that for the molecular weight we adopt and in the spin-cast type of deposition, there is no formation of crystalline P3HT domains. On the other hand, even so we see a tendency to segregation of fullerene molecules, indicating formation of nano-domains typical of bulk heterojunctions.

\section{Conclusions}

As said in the Introduction, one of the more important factors affecting efficiency in bulk heterojunction polymer/molecule photovoltaic cells is the probability of exciton migration against recombination, and the ease of charge transfer. This will be governed by the morphology of the junctions, coming from different factors including blend mass proportion, and average molecular weight of the polymer. We simulated, through our finely tuned force-field based on the UFF\cite{rapp+92jacs} and cautious CMD procedure, different blend mixing for P3HT/C$_{60}$. We find that, even with the relatively low molecular weight of $\sim$5kg$\cdot$mol$^{-1}$ where there will be no polymer crystallization, fullerene segregation is dominant. This indicates that the use of simple fullerene for the blend should bring high performance to photovoltaic cells.


\begin{acknowledgement}
The authors are grateful to M. Alves-Santos for useful discussions.
We acknowledge support from INEO, FAPESP and CNPq (Brazil).
\end{acknowledgement}
\bibliography{names,thio-uff}
\newpage

\end{document}